\documentclass[aip,preprint]{revtex4-1}
\usepackage{latexsym}        
\usepackage{graphicx}        
\usepackage{rotating}        
\usepackage{epstopdf}
\usepackage{epsfig}
\usepackage{bm}
\draft 

\begin{document}


\title{Effects, Determination, and Correction of Count Rate Nonlinearity in Multi-Channel Analog Electron Detectors}


\author{T. J. Reber}
\altaffiliation{Now at Condensed Matter Physics and Material Science Department, Brookhaven National Lab, Upton, NY, 11973}
\affiliation{Dept. of Physics, University of Colorado, Boulder, 80309-0390, USA
}%
\author{N. C. Plumb}
\altaffiliation{Now at Swiss Light Source, Paul Scherrer Institut, CH-5232 Villigen PSI, Switzerland
}%
\affiliation{Dept. of Physics, University of Colorado, Boulder, 80309-0390, USA
}%

\author{J. A. Waugh}
\affiliation{Dept. of Physics, University of Colorado, Boulder, 80309-0390, USA
}

\author{D. S. Dessau}
\affiliation{Dept. of Physics, University of Colorado, Boulder, 80309-0390, USA
}%

\date{\today}

\begin{abstract}
Detector counting rate nonlinearity, though a known problem, is commonly ignored in the analysis of angle resolved photoemission spectroscopy where modern multichannel electron detection schemes using analog intensity scales are used.  We focus on a nearly ubiquitous ``inverse saturation" nonlinearity that makes the spectra falsely sharp and beautiful. These artificially enhanced spectra limit accurate quantitative analysis of the data, leading to mistaken spectral weights, Fermi energies, and peak widths.  We present a method to rapidly detect and correct for this nonlinearity.  This algorithm could be applicable for a wide range of nonlinear systems, beyond photoemission spectroscopy.
\end{abstract}


\maketitle 

\section{Introduction}

As the technology of angle resolved photoemission spectroscopy (ARPES)\cite{KoralekLaserARPES,DamascelliRMP} continues to advance and new discoveries are made, one must be sure to eliminate all experimental artifacts from the data.  One well-known but commonly ignored effect is the detector nonlinearity.  The nonlinearity of photo-electron detectors used by Scienta, which dominates the ARPES field, was first detected by Fadley et al. during a multi-atom resonant photoemission spectroscopy (MARPES) experiment \cite{KayNonlinearity, MannellaNonlinearity,NordlundNonlinearity}.  As angle resolved photoemission spectroscopy (ARPES) is in a fundamentally different regime, such nonlinearity has generally been ignored, with a few exceptions\cite{ReberArcs, ReberPrepairing, ParhamImpurity, KordyukNL, SmallwoodScience}.  However, we find the effects of the nonlinearity, though subtle, are pernicious and must be compensated before any analysis beyond the most rudimentary can be trusted.  Here, we present the first detailed discussion of the effects of this nonlinearity as well as a new method to quickly detect and correct for this nonlinearity.
\begin{figure}[htbp]
  \begin{center}
    \leavevmode
    \includegraphics[width=140mm]{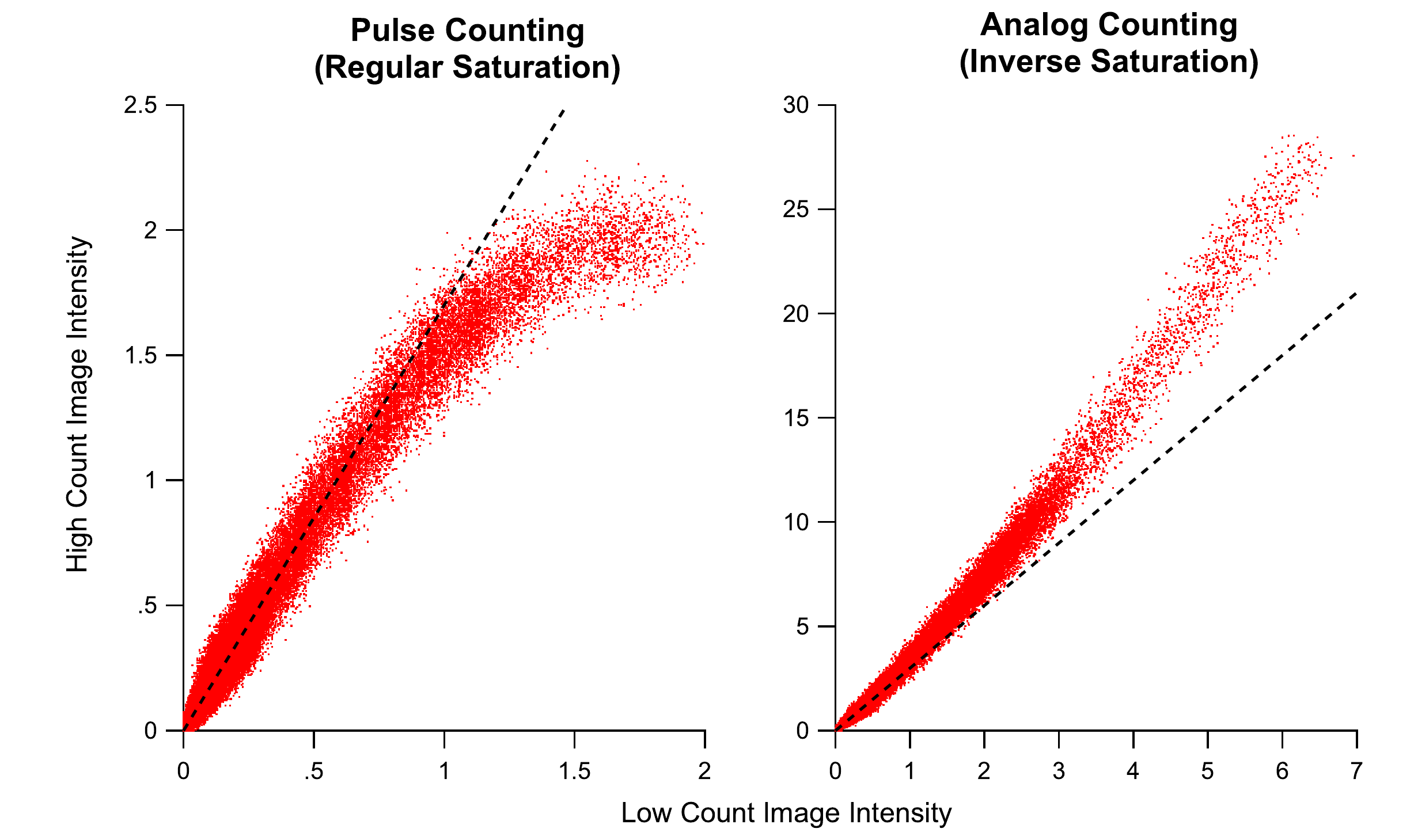}
    \caption[Comparison of ``Normal" and ``Inverse" Saturation]{{\bf Comparison of ``Normal" and ``Inverse" Saturation}.  Scatter plot of count rates per pixel of a high photon flux images (vertical) vs. low photon flux images (horizontal) (see text for an explanation of the units). Most detectors show saturation at high count rates resulting in a flattening when a high count image is plotted versus a low count image at high absolute count rates (panel a).  The analog detection schemes in modern multi-channel plate detectors have a non-linear the unusual effect of enhancement at high count rates causing a steeper slope at high count rates (panel b), i.e. an ``inverse" saturation effect. The count rate units of the two plots are in different units and so can't be directly compared.
	}
    \label{fig:NLSaturation}
  \end{center}
\end{figure}

Detecting the nonlinearity of an photo-emission spectroscopy setup involves varying the photon flux (the input) over a wide range while simultaneously plotting the photoelectron counts (the response). Such a plot will be linear if the system has a linear response\cite{KayNonlinearity}.  Such a seemingly simple test is actually difficult or impossible in most ARPES systems because the light sources that most of them utilize either can not be easily varied over a very wide range, do not have a perfectly calibrated photon counter, or both.  Even if such tools are available, this characterization is considered a time consuming endeavor and so it is rarely carried out.

Figure \ref{fig:NLSaturation} briefly presents our new technique for detecting the nonlinearity, which will work in a multichannel setup such as a camera-based MCP/phosphour screen detection setup with ``ADC or Gray Scale" analog intensity schemes for signal intensity. Two measurements of a spectrum with high dynamic range (such as a dispersive peak crossing a Fermi edge) are made back-to-back in time, with the only change being an alteration of the incident photon flux. The absolute ratio of the photon flux is not critical, though we typically use a ratio of approximately 2. We then go through each of the two images pixel-by-pixel, making a scatter plot of the count rate of each pixel on the high count image against that of the low count image. These plots would be fully linear for the ideal detector system, though as shown here the commercial systems rarely are.

We note that the count rate scales used in figure 1 are counts per binned pixel. To convert to the total flow of information onto the detector per second we multiply the average counts per pixel in the figures (of order 1-10) by the number of total binned pixels across the detector (980 in angle and 173 in energy for the plots used here) and the number of frames per second (15) to get a total information flow for these plots of order ten MHz.

Figure \ref{fig:NLSaturation}a shows a scatter plot from the pulse counting mode of a Scienta detector, showing saturation behavior at high relative count rates (upper right part of image). In practice, this saturation effect occurs at such low absolute count rates that the pulse counting mode of the Scienta systems is very rarely used. In its place the analog mode is typically used as this gives a more linear dependence in the range of count rates that are readily achievable. However, as shown in figure 1b this mode is not fully linear and in fact displays an ``inverse" saturation effect consistent with previous findings using the standard method of detecting nonlinearity (varying the photon flux over a wide dynamic range)\cite{KayNonlinearity, MannellaNonlinearity,NordlundNonlinearity}. We have observed this inverse saturation effect in at least 5 individual Scienta detectors, including SES100, SES2002, and R4000 models, as well as on a Specs Phoibos 225 spectrometer. Thus this appears to be a ubiquitous problem, likely affecting all modern camera-based ARPES setups. This effect may seem minor but can significantly alter the spectra as we will show.  Later we will use scatter plots of the type shown in figure 1 to accurately correct for the observed nonlinearity.

\section{Simulating the Effects of Nonlinearity}

\begin{figure}[htbp]
  \begin{center}
    \leavevmode
    \includegraphics[width=100mm]{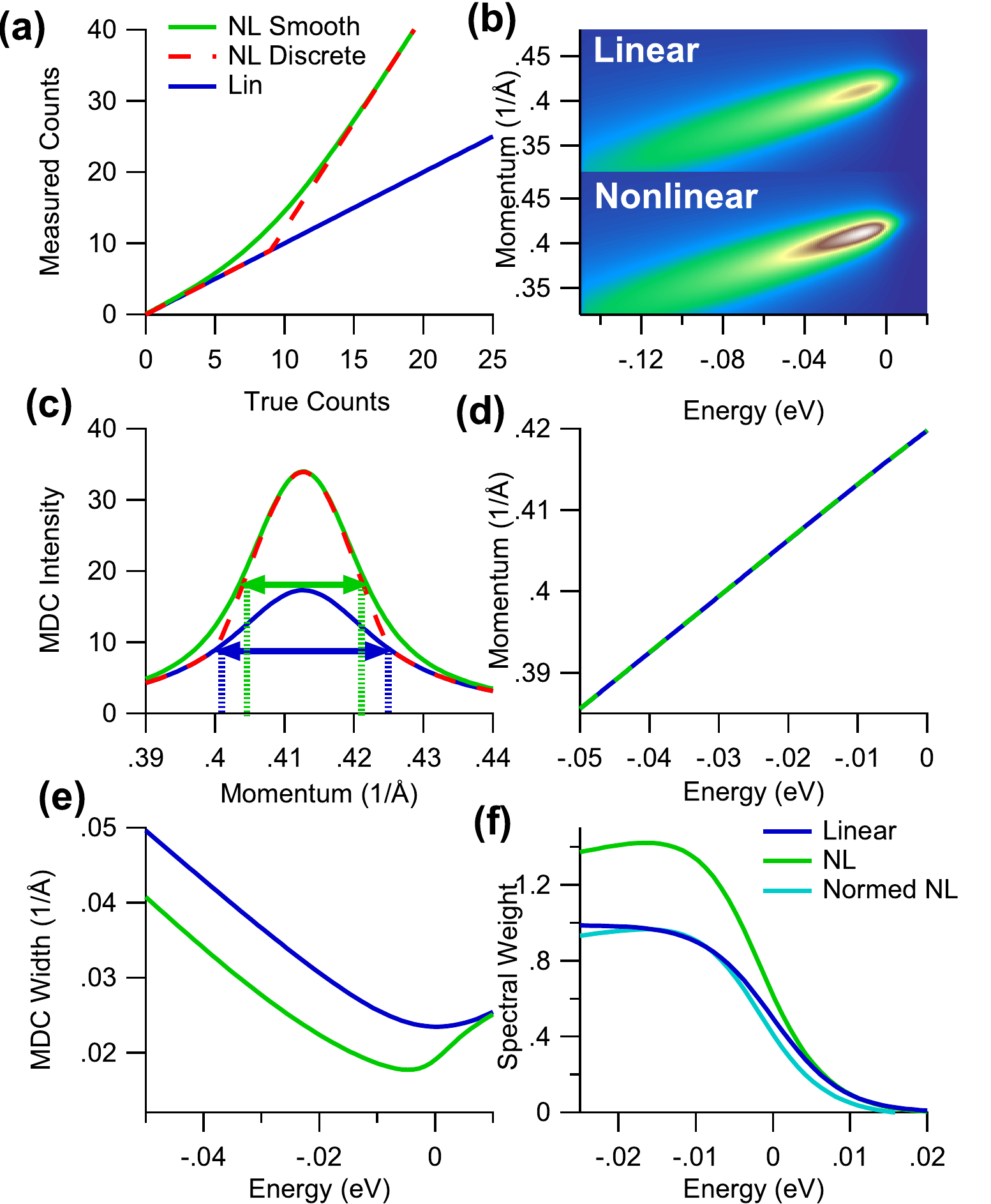}
    \caption[Effects of a Nonlinear Detector on Typical ARPES Spectra]{ {\bf Effects of a Nonlinear Detector on Typical ARPES Spectrum} {\bf (a) }Example detector nonlinearities showing both a smooth (green) and a discrete (red) deviation from a linear response (blue) {\bf (b) }Spectra before (top) and after (bottom) nonlinearity inclusion {\bf (c) }Sample MDC widths showing that the nonlinearity is one of the few experimental artifacts that make spectra sharper rather than broader  {\bf (d) }As the nonlinearity is monotonic the peaks remain the peaks, so the dispersion is unaffected by the nonlinearity  {\bf (e) } The energy dependence of the MDC widths show the narrowing expected below $E_F$, but above $E_F$ the falling spectral intensity shifts the entire MDC into the low count linear regime causing the MDC widths to return to the intrinsic value.  The resulting asymmetry in the widths should not be confused for true electron-hole asymmetry.  {\bf (f) }The spectral weights for the linear and nonlinearity spectra, showing that the asymmetric enhancement around the $E_F$ results in an apparent shifting of the Fermi energy.
	}
    \label{fig:NLBasics}
  \end{center}
\end{figure}

In Fig. \ref{fig:NLBasics}, we detail the effects of the ``inverse" nonlinearity on a simple ARPES spectrum.  We show the effect of a linear detector response (blue), and two nonlinear responses: one with a discrete change in slope (red), which makes the effects more obvious and a smoothly varying one (green) which makes the effects less obvious but is closer to what is observed.  In \ref{fig:NLBasics}a we show the nonlinearity of the detector in measured counts vs true counts (as will be shown later this is not exactly the same as the high count vs low count image).  To elucidate how the actual spectra are affected, we depict the two simple ARPES spectra, a linear one and a continuously nonlinear one, side by side in figure \ref{fig:NLBasics}b.  We assumed a linear bare band and Marginal Fermi liquid peak broadening  appropriate for near-optimally doped cuprate samples \cite{VarmaMFL,VarmaPNAS}.  In \ref{fig:NLBasics}c we show the effects of the nonlinearity on a sample momentum distribution curve(MDC)\cite{VallaMoly}.  While the deviation from a Lorentzian is obvious in the discrete case, the smoothed one is decently well described by a Lorentzian.  Consequently, detecting nonlinearity from a line-shape is difficult.   The peak of the Lorentzian does not shift when the nonlinearity is applied, so analysis based on peak locations (e.g. band mapping, dispersions, Re($\Sigma$)) are robust against the nonlinearity (\ref{fig:NLBasics}d). In the case of an asymmetric peak in momentum or two overlapping peaks, extracted peak positions could clearly be impacted by the nonlinearities.  The peak enhancement also raises the half max level, effectively narrowing the peak width. Consequently, these widths, a common measure of the electron scattering rate, can be significantly sharpened by the detector nonlinearity (\ref{fig:NLBasics}e).  However as the intensity above $E_F$ is rapidly suppressed by the Fermi edge, the distorted nonlinear widths quickly return to the linear values.  This creates a noticeable asymmetry in the widths that is roughly centered at $E_F$, which could be incorrectly interpreted as electron-hole asymmetry.  Finally, the spectral weight, determined by integrating the MDC's shows a clear enhancement due to the nonlinearity (Fig. \ref{fig:NLBasics}f).    However, with no reference this enhancement can be hard to detect in a single spectrum.

\begin{figure}[htbp]
  \begin{center}
    \leavevmode
    \includegraphics[width=120mm]{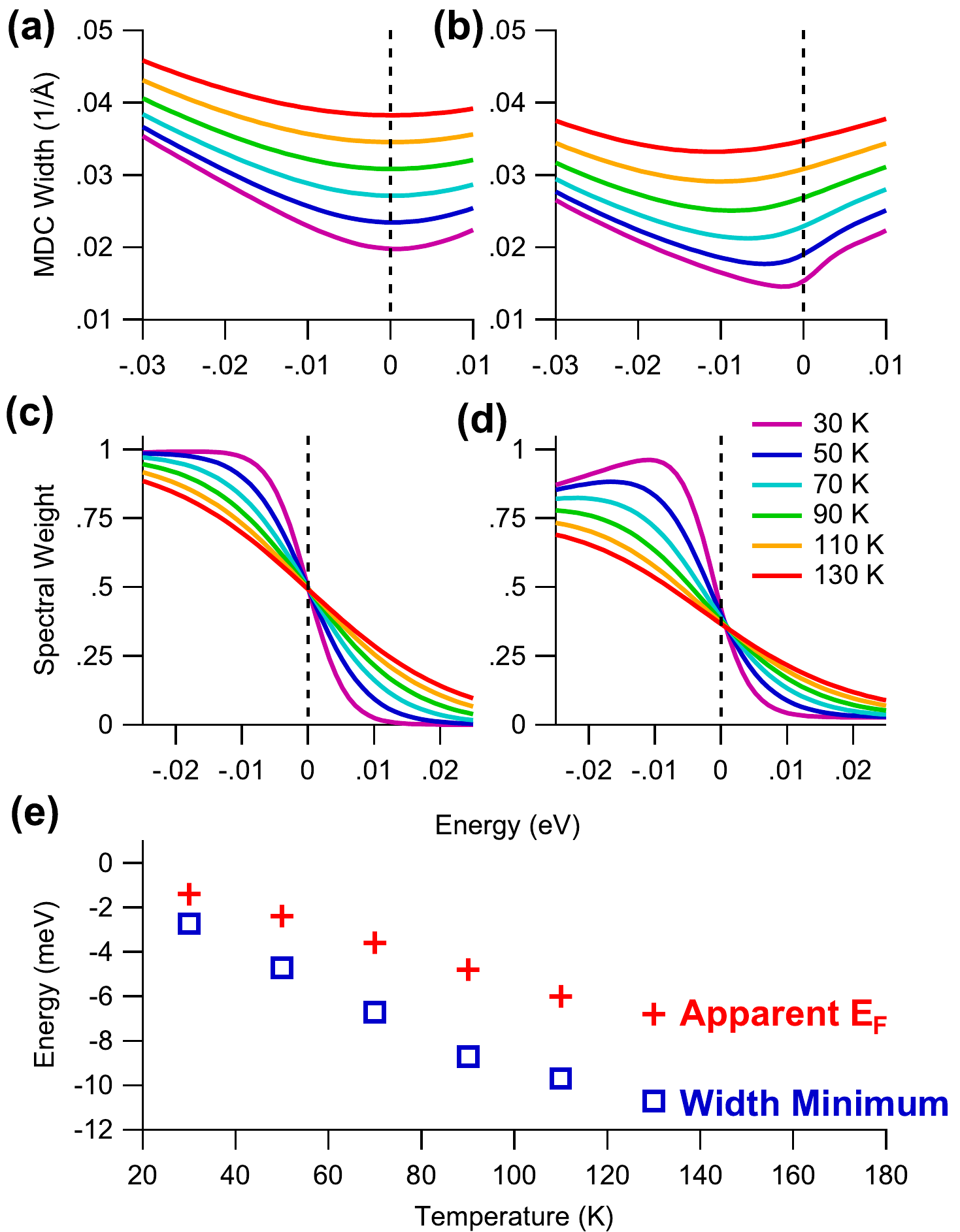}
    \caption[Effects on Nonlinearity on Temperature Dependence Studies]{ {\bf Effects on Nonlinearity on Temperature Dependence Studies} {\bf (a) }Example temperature dependence of MDC widths   {\bf (b) }Temperature dependence of widths after addition of nonlinearity, showing formation of asymmetry and shifting minimum width
    {\bf (c) }Example of temperature dependence of spectral weight with isosbestic point centered at $E_F$ and half filling  {\bf (d) }Temperature dependence of spectral weights after addition of nonlinearity showing the isosbestic point holds at $E_F$ though shifted away from half filling  {\bf (e) }Temperature dependence of the width minimum and the apparent Fermi edge location after addition of nonlinearity
	}
    \label{fig:NLTDependence}
  \end{center}
\end{figure}

Since the asymmetry and $E_F$ drift is an effect of the spectral intensity change at the Fermi edge, it is strongly temperature dependent.  To illustrate this behavior, we show a temperature dependence of the widths for a simulation of Marginal Fermi Liquid (hyperbolic energy dependence\cite{VarmaPNAS}) in figure \ref{fig:NLTDependence}a and the corresponding nonlinear ones in \ref{fig:NLTDependence}b.  Note that the asymmetry is strongest in the coldest sample but the other effect of the nonlinearity is a softening (shifting to higher binding energy) of the width minimum with decreasing temperature.  This softening is unphysical in that the minimum of the scattering rate should be pegged to $E_F$, which can be understood by considering that the allowable phase space for decay channels is minimized at the $E_F$.  This softening is more easily observed than the asymmetry so it is a clear sign of nonlinearity in the spectra.

The temperature dependence of the linear and nonlinear spectral weights show another symptom of nonlinearity (Fig.  \ref{fig:NLTDependence}c and \ref{fig:NLTDependence}d).  Namely the isosbestic point (point of constant spectral weight) for the linear term is centered at $E_F$ in energy and half filling in weight as expected for particle conservation.  However in the nonlinear case the isosbestic point deviates slightly from $E_F$ (here it is a very subtle effect) and its filling is less than half of the max value so apparent particle conservation is broken. For the simulations already described we show the temperature dependence of both in  Fig. \ref{fig:NLTDependence}e. If these edges were utilized to determine the experimental $E_F$ we would obtain the false appearance of a temperature dependent $E_F$ and minimum width location. Worse, if the reference spectra (say a polycrystalline Au) had a different count rate than the sample that science was being carried out on (say a superconductor whose gap was being measured), then each spectrum would have different shifts from the true Fermi edge location. Such a drifting Fermi edge calibration would have deleterious effects on procedures which require highly accurate determination of the experimental $E_F$, particularly gap measurements and spectra where the Fermi function is divided out in order to extract information about thermally occupied states above $E_F$.

\section{Correcting for the Nonlinearity}
Now that we have discussed a method to detect the nonlinearity as well as its many effects on the measured spectra, we here discuss a method to process the data so as to remove the major effects of the nonlinearity.  This technique has two implicit assumptions.  First, we assume the nonlinearity is uniform across the detector, which is reasonable as the standard method of taking data in ARPES involves sweeping the spectrum across the entire detector effectively averaging out any inhomogeneity.   Second, we assume that the very lowest counts region is representative of the true counts, which is justified as the slope of the high count vs low count is comparable to the change in the  photon flux.

We begin with the high count vs. low count scatter plots such as those shown in figure 1. While these are not the actual nonlinearity curves (measured counts vs. true counts) they do contain all the information necessary to extract the nonlinearity correction. To remove the statistical spread, we fit the high count vs. low count plot (red in Fig. 4b)  with a high-order monotonic polynomial fit (green) from which the nonlinear correction will be extracted.

\begin{figure}[htbp]
  \begin{center}
    \leavevmode
    \includegraphics[width=160mm]{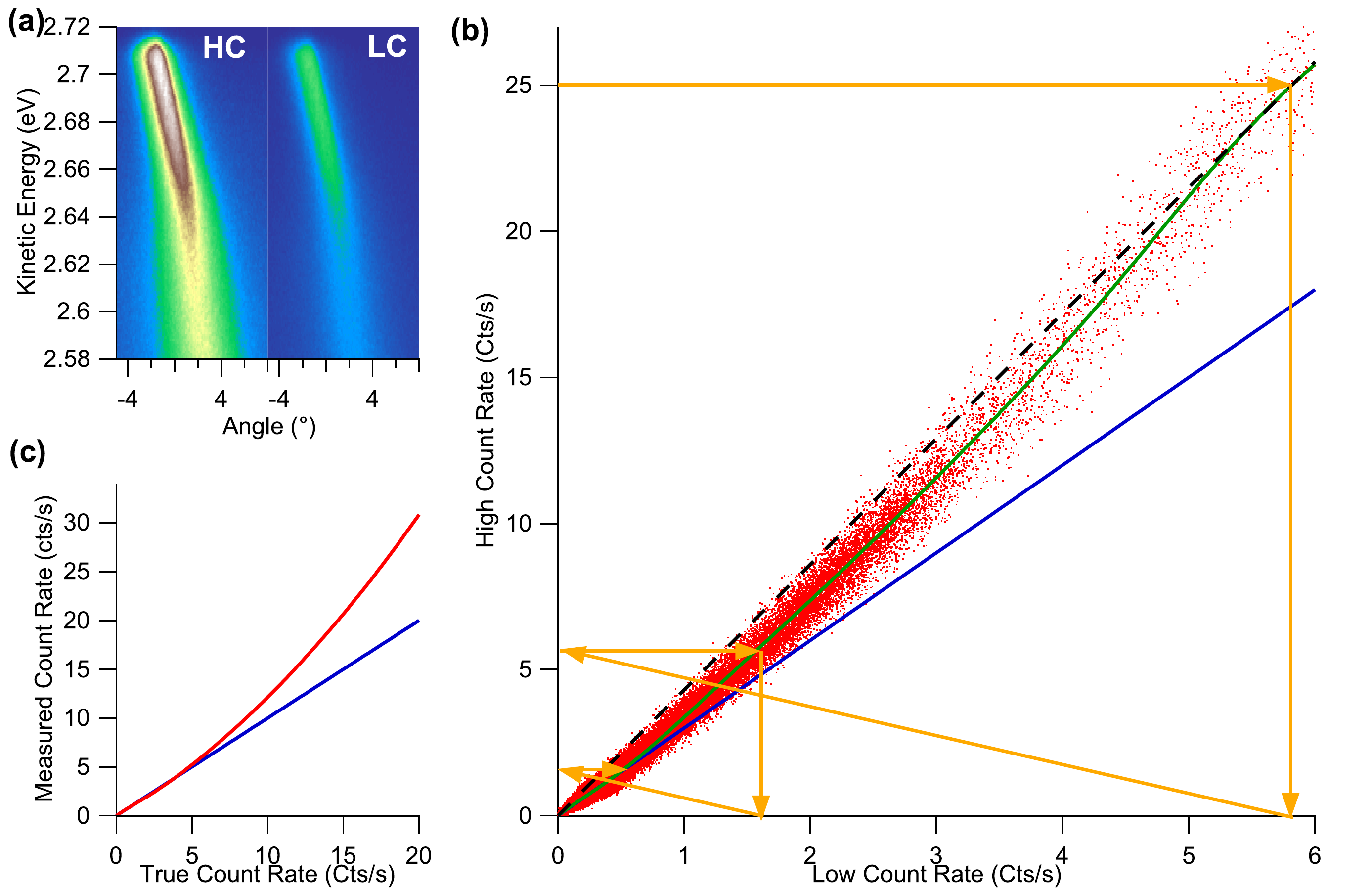}
    \caption[Extracting the Nonlinearity]{ {\bf Extracting the Nonlinearity} {\bf (a) }High count image and low count image  {\bf (b) }High Count vs Low Count scatter plot (red), high order polynomial fit (green) and low count linear extrapolation (blue) and gold arrows tracing the nonlinear extraction method's iterations. {\bf (c) } Extracted nonlinear curve (red) and low count linear extrapolation (blue).  A similar plot as panel c has been obtained by varying the photon flux over a wide dynamic range \cite{KayNonlinearity}.
	}
    \label{fig:NLextraction}
  \end{center}
\end{figure}

The algorithm to extract the nonlinearity is composed of two steps which allow us to first iteratively reach the linear low count regime and then extrapolate back to the underlying true counts. The method is shown schematically in fig 4b.  We start with a given point on the green fit and determine the ratio of measured high counts to the measured low counts, knowing that the actual change in the true counts is the ratio of photon fluxes.  Then we shift down the green curve (following the gold arrows) until the high counts now equal the old low count value and again find the ratio of high counts to low counts for that new point.  This process is iterated until we enter the linear regime. In the linear regime, the measured counts are the true counts, and we know the number of iterations and thus the number of flux ratios we traverse, so it is simple extrapolation back up to find the underlying true counts for the original high count value.  We repeat the process for every high count value and we can build up the detector's nonlinear response curve (red in Fig. \ref{fig:NLextraction}c).  The response clearly deviates from linear (blue).

The nonlinearity extraction algorithm is shown in the next few lines.
\begin{equation} \label{eq:NLExtract1}
\Omega=\frac{HC(x_1)}{LC(x_1)}\frac{HC(x_2)}{LC(x_2)}\frac{HC(x_3)}{LC(x_3)}\cdot\cdot\cdot\frac{HC(x_n)}{LC(x_n)}
\end{equation}
Which if we express in terms of the nonlinearity function acting on the original true count rate at the $x_1$.
\begin{equation} \label{eq:NLExtract2}
\Omega=\frac{NL(I)}{NL(I/R_F)}\frac{NL(I/R_F)}{NL(I/R_F^2)}\frac{NL(I/R_F^2)}{NL(I/R_F^3)}\cdot\cdot\cdot\frac{NL(I/R_F^{n-1})}{NL(I/R_F^{n})}
\end{equation}
This can be simplified to:
\begin{equation} \label{eq:NLExtract3}
\Omega=\frac{NL(I)}{NL(I/R_F^n)}
\end{equation}
Since we stop the iteration in the linear regime
\begin{equation} \label{eq:NLExtract4}
NL(I/R_F^n)=I/R_F^n
\end{equation}
which can be simplified to:
\begin{equation} \label{eq:NLExtract5}
\Omega=\frac{NL(I)}{I/R_F^n}
\end{equation}
Since we know the values of $\Omega$, $NL(I)$, $R_F$ and $n$, it is simple to extract $I$.  Repeating this procedure for each point on the  high count vs low count fit, we can extract the full nonlinear curve.  This method is more general than that proposed by Kordyuk et al. \cite{KordyukNL} as it does not presume a form for the nonlinearity.  In fact this algorithm is general enough to be used in fields outside of ARPES that have uniform nonlinearity across a two-dimensional detector.

This algorithm does fail when the assumption of linearity in the low count region is not valid.  For instance, if the detector had a quadratic response with no linear dependence then the HC/LC ratio would be linear even though the response is not.  For an arbitrary power $n$:
\begin{equation} \label{eq:AlFail1}
HC=NL(I)=I^n
\end{equation}
and
\begin{equation} \label{eq:AlFail2}
LC=NL(\frac{I}{R_F})=\frac{I^n}{R_F^n}
\end{equation}
So
\begin{equation} \label{eq:AlFail3}
HC(LC)=R_F^n*LC
\end{equation}
Consequently, while the high count vs low count curve may appear linear the slope reveals if the low count linearity assumption is valid or not.  For the detectors we've studied that assumption is valid.  As the extracted nonlinearity from this method closely matches that measured by the much more laborious flux variation method\cite{KayNonlinearity,MannellaNonlinearity}, we do not expect it to be an inherent error of this new extraction method.

Because of the proprietary nature of the detection schemes used in these analyzers it is hard to exactly determine the origin of the observed nonlinearity. However, a few likely candidates exist.  First, phosphor has a well known ``inverse saturation" type nonlinearity with kinetic energy of impacting electrons (necessitating the gamma correction on cathode ray tubes.) It is not unreasonable that the phosphor might have a nonlinear response to the electron flux as well.  Second, the background subtraction or thresholding must be done to remove the very low signal strengths associated with electronic noise or camera read-out noise - a problem that is compounded by the widely varying signal strengths per event coming out of the micro-channel plates. If the thresholding is too aggressive, larger fractions of signal would be removed from the low count regions than the high count regions, creating a nonlinear response. Third, the output data from these systems undergoes significant proprietary processing with the built-in software and firmware.  This nonlinearity could be an unforseen consequence of that processing.

\section{Testing the Corrected Data}

\begin{figure}[htbp]
  \begin{center}
    \leavevmode
    \includegraphics[width=160mm]{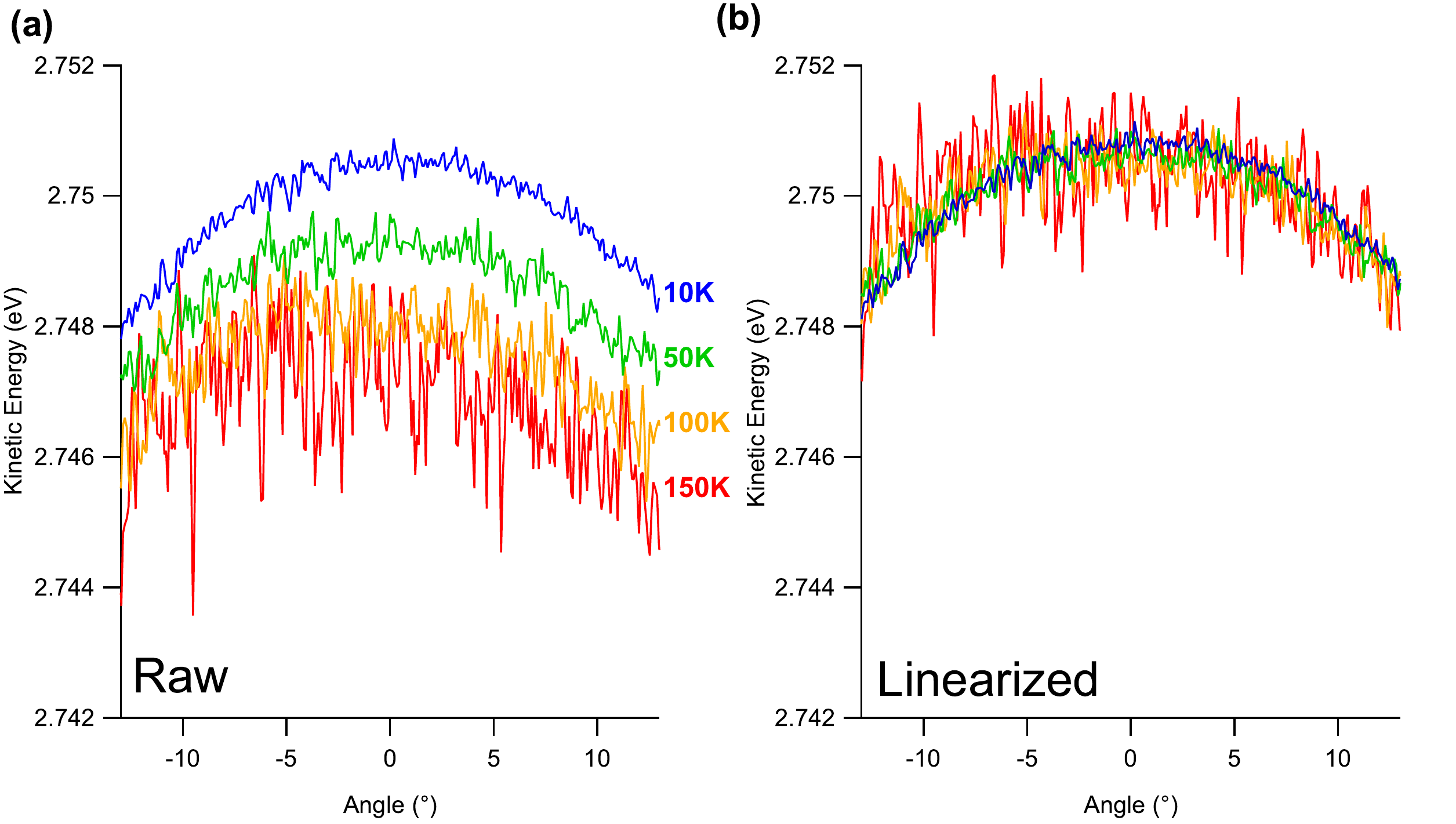}
    \caption[Linearizing Amorphous Gold]{ {\bf Linearizing Amorphous Gold}  The shifting Fermi edge with temperature from nonlinearity is evident in amorphous gold and can be corrected with the nonlinearity extraction. The curvature in angle is a known effect of straight slits at the entrance of the curved hemispherical analyzer, and is readily corrected.
	}
    \label{fig:GoldNL}
  \end{center}
\end{figure}

One of the simpler tests for the detector nonlinearity is the temperature dependence of an amorphous gold sample.  Amorphous (non-crystalline) gold is an ideal reference when taking ARPES data.  The non-reactive nature of gold makes it resistant to aging, and the amorphous nature averages over all the bands such that the spectra are uniform in angle but still show the Fermi edge at $E_F$.  Consequently, gold is regularly used to correct for detector inhomogeneity, as well as to empirically determine both the Fermi energy as well as the resolution of the instrument.  Even this simplest of ARPES data manifests the shifting Fermi edges due to the nonlinearity, but after correction with the curve extracted from Bi2212 spectra the Fermi edges no longer show any sort of thermal drift as expected (Fig. \ref{fig:GoldNL}).

\begin{figure}[htbp]
  \begin{center}
    \leavevmode
    \includegraphics[width=160mm]{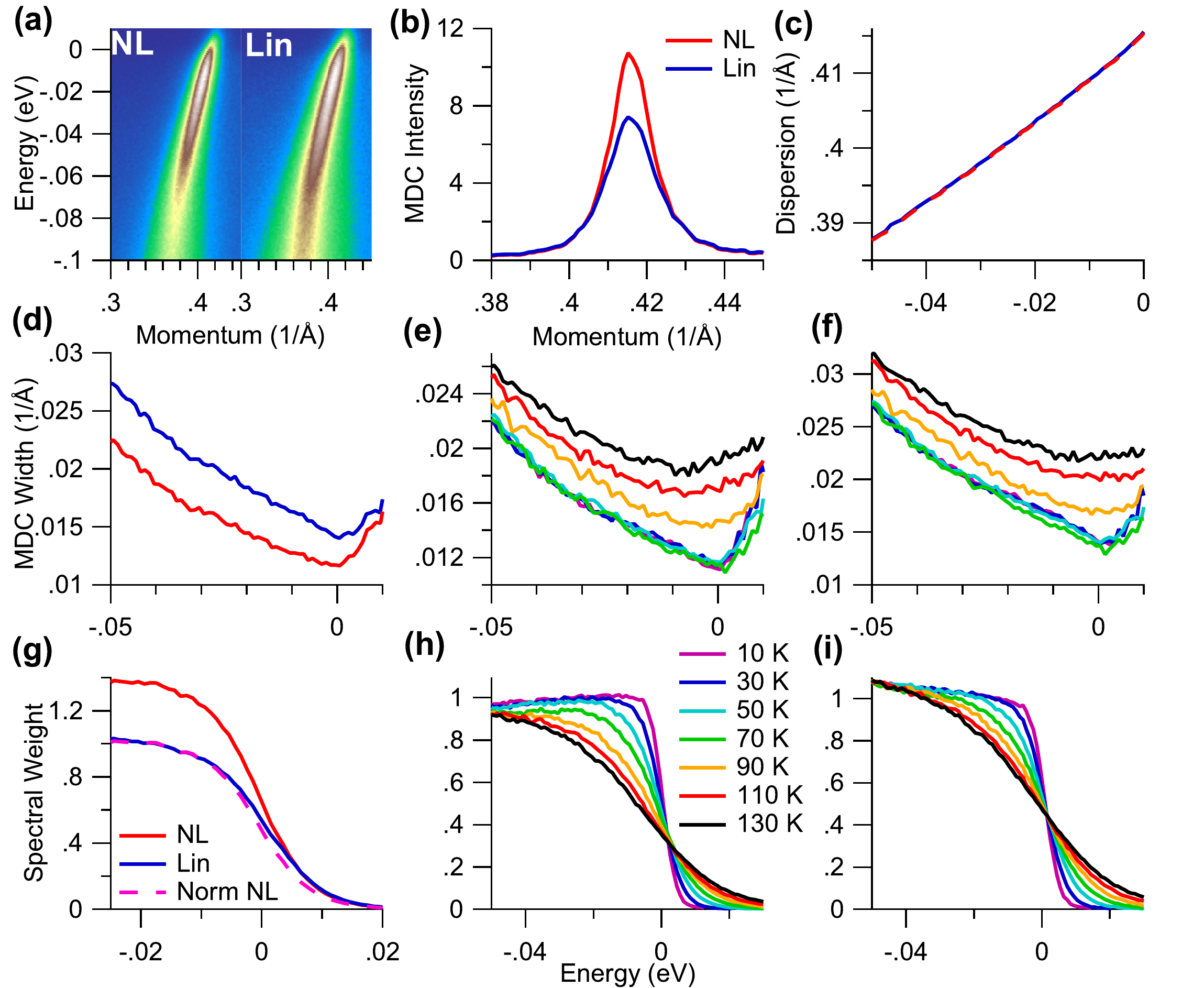}
    \caption[Effects of Linearizing Data]{ {\bf Effects of Linearizing Data}  {\bf (a) } Example nodal spectra of Bi$_2$Sr$_2$CaCu$_2$O$_{8+\delta}$ before and after linearization {\bf (b) }Effects of linearization on sample MDC {\bf (c) }Effects of linearization on dispersion {\bf (d) }Effects of linearization on MDC widths {\bf (e) }Temperature dependence of raw MDC widths {\bf (f) } Temperature dependence of linearized MDC widths  {\bf (g) }Effects of linearization on spectral weight   {\bf (h) } Temperature dependence of raw spectral weights with isosbestic point well below half filling {\bf (i)} Linearized spectral weights
	}
    \label{fig:LinearizingData}
  \end{center}
\end{figure}

Furthermore, we show on experimental data the difference between nonlinear and linearized Bi2212 results (Fig. \ref{fig:LinearizingData}), showing many unusual features: drifting minimum widths, electron-hole asymmetric widths, low isosbestic points, are all absent or significantly less pronounced in the linearized data.  The remnant oddities are likely due to an imperfect linearization rather than representative of true features.

\section{Conclusion}

While the effects of nonlinearity are greatly mitigated by the procedure outlined above, it is impossible to be completely certain that the nonlinearity is fully removed for the very low count rate portions of the spectra, which is where the nonlinearity comes into play for the analog counting modes (fig 1b). The best option to ensure full-linearity for the low count portions is to utilize a pulse counting scheme (fig 1a), except that presently available commercial schemes for this then suffer from nonlinearity at higher count rates. In this regard it is helpful to note that as long as the ``regular" saturation is not too severe the scheme presented here can also be used to correct for this form of saturation.

We have presented a detailed study of the effects of the typical detector counting rate nonlinearity on a simple ARPES spectrum.  While studies that have focused on peak positions are almost fully unaffected by this experimental artifact, studies of the peak widths and spectral weight can be significantly distorted.  Additionally, any report whose finding is critically sensitive to the accurate determination of $E_F$ could be negatively influenced by this detector nonlinearity. We also present a simple method to rapidly detect and then largely correct for this counting rate nonlinearity.


%
%

\begin{acknowledgments}
We thank D. H. Lu and R. G. Moore for help at SSRL and M. Arita and H. Iwasawa at HiSOR. SSRL is operated by the DOE, Office of Basic Energy Sciences.  Funding for this research was provided by DOE Grant No. DE-FG02-03ER46066.

\end{acknowledgments}

\end{document}